\documentclass[twocolumn,showpacs,preprintnumbers,amsmath,amssymb]{revtex4}
\usepackage{dcolumn}
\usepackage{bm}
\usepackage{graphicx}

\begin{document}
\title{Physical solution of 1D quantum $N$-boson system with zero-range pair interaction}
\author{Wenhua Hai, \  Guishu Chong, \ Qiongtao Xie} \affiliation{
Department of physics, Hunan Normal University, Changsha 410081,
China}


\email{adcve@public.cs.hn.cn}
\begin{abstract}
The mathematically exact solution of a one-dimensional (1D)
quantum $N$-identical-boson system with zero-range pair
interaction has been well known. We find that this solution is
non-physical, since there exists a paradox of its energy
expectation value, leading a basic contradiction to the mean-field
theory of the system. A new integral equation that is equivalent
to the corresponding Schr\"{o}dinger one is established and its
formally physical solution is derived. Energy correction from the
average potential and harmonic waves of different momentums are
demonstrated, and scattering amplitude of the physical solution
and mean-field theory of the system are discussed.
\end{abstract}

\pacs{03.65.Ge; 03.65.Nk; 05.30.Jp; 11.80.Fv}

\maketitle

The quantum integrability is an essential property of quantum
mechanics and field theory. A kind of non-trivial integrable
Hamiltonian operators, which describes some systems consisting of
$N$ identical non-relativistic particles interacting pairwise
through some forms of potentials has been generally demonstrated
by using different methods \cite{Polychronakos}-\cite{Mattis}. The
forms of potentials contain the zero-range infinite pair
interaction \cite{Lee}-\cite{Yang} and the long-range finite pair
interactions \cite{Calogero}-\cite{Ullate}. Although the
integrable models may not be exactly solvable, the exactly
solvable models are included by them. Much interest have been
focused on the exactly solvable models, because of their
applications to the Bose-Einstein condensations
\cite{Lieb}-\cite{Leggett}, quantum Yang-Baxter equation
\cite{Ge2}, quantum spin chains \cite{Haldane},
\cite{Polychronakos2}, random matrix theory \cite{Simons},
fractional statistics \cite{Ha}, \cite{Isakov}, Yang-Mills
theories \cite{Minahan}, \cite{Hoker}, quantum Hall liquids
\cite{Azuma}, \cite{Iso}, soliton theory \cite{Polychronakos3},
and the black holes \cite{Gibbons}.

The infinite zero-range pair interactions can be expressed as the
$\delta$ function potentials, which approximate some practical
short-range interactions and seem to be easier for mathematical
treatment than the latter. However, singularity of the $\delta$
function may bring us some new troubles that lead to non-physical
results. This can be shown through the simplest example, the
Schr\"{o}dinger equation $H\psi=- \frac 12 \psi _{x_r x_r} + c
\delta (x_r) \psi =E\psi$ in atomic unit and with the $\delta$
function potential $c \delta (x_r)$, where $x_r$ denotes the
relative coordinate of two particles, $\psi(x_r)$ represents the
relative wave function, $c>0$ is the zero-range interaction
intensity and $E$ the eigenenergy of the Hamiltonian $H$. The
standard quantum mechanical book \cite{Zeng} had mathematically
given its general solution as the linear combination of plane
waves, $\psi =ae^{ikx_r}+be^{-ikx_r}$ with $k=\sqrt{2E}$ and $a, \
b$ being constants determined by the normalization and boundary
conditions. This result had also been extrapolated to $N$-body
zero-range pair interaction problems \cite{Ge2}-\cite{Lieb} and
been recently applied to few-particle Bose-Einstein condensates
\cite{Cirone}. However, there is a paradox of the energy
expectation value in this mathematical solution. Noticing that the
function $e^{\pm ikx_r}$ denotes the state of a free particle, in
this state the average kinetic energy is equal to the total energy
and the average potential is $\langle \psi |c \delta (x_r)|\psi
\rangle=c|\psi (0)|^2$. The nonzero average potential leads to the
paradox of average energy, namely the energy value may be $\langle
\psi | H|\psi \rangle=E$ or $\langle \psi | H|\psi
\rangle=E+c|\psi(0)|^2$. Both results are in agreement only for
the identical-fermion system with $\psi(0) = 0$. While for an
$N$-identical-boson system with $\psi(0)\ne 0$ the mean-filed
theory of second quantization is based on the average Hamiltonian
and therefore there exists basic contradiction from the above
result. Although adopting the finite short-range interaction
\cite{Girardeau}, \cite{Blume}-\cite{Bedaque} instead of the
infinite zero-range interaction can avoid the physical confusion,
the Gross-Pitaevskii's nonlinear Schr\"{o}dinger equation comes
from the application of the zero-range interaction, which has been
successfully used to treat the Bose-Einstein condensates
\cite{Dalfovo}-\cite{Leggett}. This means existence of the
physical solution of 1D $N$-boson system with zero-range pair
interaction.

In the previous works \cite{Hai}, \cite{Hai2}, we established and
employed an integral equation to find the complete information
implied in Schr\"{o}dinger's quantum systems. Applying this idea
to write down an new integral equation of the zero-range pair
interacting system and seeking its formally physical solution to
eliminate the contradiction on energy expectation value are our
main motive in this paper. In the physical states, the energy
correction from the pair interaction and harmonic waves with
different momentums are investigated, and the scattering amplitude
and self-consistent field theory of the system are discussed.

We commence with $N$ slow identical bosons propagating in $x$
direction and assume that the pairwise interaction consists only
of a hard core of 1D diameter. For small diameter the system is
conveniently treated as impenetrable point particles with the
delta-function pseudopotential $c\delta (x_j-x_i)$, where $c>0$ is
the interaction strength, $x_i, \ x_j$ for $i, j=1, 2, \cdots, N$
are the coordinates of particles $i$ and $j$. The stationary state
Schr\"{o}dinger equation pertinent to the system reads as
\cite{Ge2}-\cite{Yang}
\begin{eqnarray}
-\sum_{i=1}^N \frac{\hbar^2}{2m} \bigtriangledown ^2 \psi
+c\sum_{i\ne j}\delta (x_j-x_i)\psi =E \psi
\end{eqnarray}
with $m$ and $E$ being the mass of a single particle and the
energy of the whole system respectively. Save for the collisions,
which occur only for $x_i=x_j$, the previous works treated the
system as $N$ free particles and constructed its exact plane wave
solution mathematically \cite{Ge2}-\cite{Yang}. Obviously, the
above-mentioned paradox of energy expectation value exists in this
solution. In order to seek the physical solution without the
paradox, we must count in energy correction from the pair
interaction  and solve this equation by using a new approach.

Under assumption of the pairwise interaction and impenetrability
of particles, the exchange symmetric wave function can be
expressed as \cite{Ge2}, \cite{Mattis}
\begin{eqnarray}
\psi &=& \sum_p \psi (x_{p_1}\cdots x_{p_N})\prod_{odd \ i} \theta
(x_{p_{i+1}}-x_{p_i}) \nonumber \\ &=&  \sum_p \prod_{odd \ i}
\phi(x_{p_{i+1}}, \ x_{p_i})\theta (x_{p_{i+1}}-x_{p_i}).
\end{eqnarray}
Here $x_{p_1}\cdots x_{p_N}$ denotes a permutation of $x_1 \cdots
x_N$, $\sum_p$ represents sum of all the permutations,
$\theta(x_{p_{i+1}}-x_{p_i})$ is the step function satisfying
$\theta(x_{p_{i+1}}-x_{p_i})=1$ for $x_{p_{i+1}}>x_{p_i}$ and
$\theta(x_{p_{i+1}}-x_{p_i})=0$ for $x_{p_{i+1}}< x_{p_i}$, and
$\phi(x_{p_{i+1}}, \ x_{p_i})$ is the wave function of two
particles. To form whole particle pairs, we take the large number
$N$ as an even number and approximate
\cite{Dalfovo}-\cite{Leggett} $N \pm 1$ to the even $N$. The
exchange symmetry of the identical bosons requires $\phi(x_{p_i},
\ x_{p_{i+1}})=\phi(x_{p_{i+1}}, \ x_{p_i})$. Substituting Eq. (2)
into Eq. (1), we get the Schr\"{o}dinger equation of two-boson
wave function \cite{Ge2}, \cite{Cirone}
\begin{eqnarray}
-\sum_{j=i}^{i+1} \frac{\hbar^2}{2m} \frac {\partial ^2
\phi}{\partial x^2_{p_j}} +c\delta (x_{p_{i+1}}-x_{p_{i}})\phi
=E_{i} \phi,
\end{eqnarray}
where $E_{i}$ is the energy of two particles, obeying $\sum_i
E_i=E$ for $i=1,3,5,\cdots,N-1$. Introducing the mass-center and
relative coordinates
\begin{eqnarray}
x_{ic}=\frac {1}{\sqrt{2}}(x_{p_{i+1}}+x_{p_{i}}), \ \ \
x_{ir}=\frac {1}{\sqrt{2}}(x_{p_{i+1}}-x_{p_{i}})
\end{eqnarray}
to Eq. (3) and setting
\begin{eqnarray}
\phi(x_{p_{i+1}}, \
x_{p_i})=\phi_{ic}^{(p)}(x_{ic})\phi_{ir}^{(p)}(x_{ir}), \ \ \
E_{i}=E_{ic}+E_{ir},
\end{eqnarray}
we obtain the equation of relative motion
\begin{eqnarray}
-\frac{\hbar^2}{2m} \frac {d^2 \phi_{ir}^{(p)}}{d x^2_{ir}}
+\frac{c}{\sqrt{2}}\delta (x_{ir})\phi_{ir}^{(p)} =E_{ir}
\phi_{ir}^{(p)}
\end{eqnarray}
and the mass-center motion equation. In the calculation, the
formula $\delta(x_{p_{i+1}}-x_{p_{i}})
=\delta(\sqrt{2}x_{ir})=\delta(x_{ir})/\sqrt{2}$ has been
employed. The mass-center equation is that of a free particle with
eiginenergy $E_{ic}$ and eiginstate
$\phi_{ic}^{(p)}=a_ie^{ik_{ic}x_{ic}} +b_ie^{-ik_{ic}x_{ic}}$,
where $a_i$ and $b_i$ are undetermined constants and
$k_{ic}=\sqrt{2mE_{ic}/\hbar^2}$.

The corrected energy from the delta potential is of the order
$c|\phi_{ir}^{(p)}(0)|^2$, which is set as $E'_{ir}$, so we have
\begin{eqnarray}
E_{ir}&=& E_{ir}^{(0)}+E'_{ir}, \ k^2_{ir}= 2mE_{ir}/\hbar^2
=k_{ir}^{2(0)}+k'^2_{ir}, \nonumber
 \\
k_{ir}^{(0)}&=& \sqrt{\frac{2mE_{ir}^{(0)}}{\hbar^2}}, \
k'_{ir}=\sqrt{\frac{2mE'_{ir}}{\hbar^2}}, \
c'=\frac{\sqrt{2}mc}{\hbar^2}.
\end{eqnarray}
The corresponding momentum reads as $\hbar k_{ir}\approx \hbar
[k^{(0)}_{ir}+k'^2_{ir}/(2k^{(0)}_{ir})]$. Inserting them into Eq.
(6) yields a simplified equation, then following Ref. \cite{Hai}
leads to the new integral equation
\begin{eqnarray}
\phi_{ir}^{(p)}&=& Ae^{ik_{ir}^{(0)}x_{ir}}+Be^{-ik_{ir}^{(0)}x_{ir}}- \frac{i}{2k_{ir}^{(0)}} \nonumber \\
&\times&
\Big\{e^{ik_{ir}^{(0)}x_{ir}}\int_{C}e^{-ik_{ir}^{(0)}x_{ir}}[c'\delta(x_{ir})-k'^2_{ir}]\phi_{ir}^{(p)}dx_{ir}
\nonumber \\
&-&e^{-ik_{ir}^{(0)} x_{ir}}\int_{D}e^{ik_{ir}^{(0)}
x_{ir}}[c'\delta(x_{ir})-k'^2_{ir}]\phi_{ir}^{(p)}dx_{ir}\Big\},
\nonumber \\
\end{eqnarray}
where $A, \ B, \ C$ and $D$ are undetermined constants obeying
$C<0, \ D<0$. Applying Eq. (8) to Eq. (6), we can directly prove
the agreement between them. Particularly, the relative wave
function $\phi_{ir}^{(p)}$ in Eq. (8) does not satisfy the free
particle equation such that the above paradox of energy
expectation value is avoided. Completing the integrations in Eq.
(8), we get its formally exact solution as
\begin{widetext}
\begin{equation}
\phi_{ir}^{(p)}=\theta
(x_{ir})\Big\{\Big[A+\frac{c'\phi_{ir}^{(p)}(0)}{2ik_{ir}^{(0)}}
+\frac{ik'^2_{ir}}{2k_{ir}^{(0)}}\int_{C}e^{-ik_{ir}^{(0)}x_{i}r}\phi_{ir}^{(p)}dx_{ir}
\Big]e^{ik_{ir}^{(0)}x_{ir}}
+\Big[B-\frac{c'\phi_{ir}^{(p)}(0)}{2ik_{ir}^{(0)}}
-\frac{ik'^2_{ir}}{2k_{ir}^{(0)}}\int_{D}e^{ik_{ir}^{(0)}x_{ir}}\phi_{ir}^{(p)}dx_{ir}\Big]e^{-ik_{ir}^{(0)}x_{ir}}\Big\}
\nonumber
\end{equation}
\begin{equation}
+\theta
(-x_{ir})\Big\{\Big[A+\frac{ik'^2_{ir}}{2k_{ir}^{(0)}}\int_{C}e^{-ik_{ir}^{(0)}x_{ir}}\phi_{ir}^{(p)}dx_{ir}
\Big]e^{ik_{ir}^{(0)}x_{ir}} +
\Big[B-\frac{ik'^2_{ir}}{2k_{ir}^{(0)}}\int_{D}e^{ik_{ir}^{(0)}x_{ir}}\phi_{ir}^{(p)}dx_{ir}\Big]e^{-ik_{ir}^{(0)}x_{ir}}\Big\},
\end{equation}
\end{widetext}
where $\phi_{ir}^{(p)}(0)$ is the value of $\phi_{ir}^{(p)}$ at
$x_{ir}=0$. Applying Eq. (9) and the above-mentioned
$\phi_{ic}^{(p)}$ to Eq. (5), then to Eq. (2), we obtain the
formally exact solution of the $N$-identical-boson system. The
formal solution, of course, is still an integral equation, since
$\phi_{ir}^{(p)}$ appears in the integrations of Eq. (9). It is
worth noting that the integral equation (9) completely describes
the continuity of $\phi_{ir}^{(p)}$ and discontinuity of
$d\phi_{ir}^{(p)}/dx_{ir}$ caused by the singularity of delta
potential.

When the constants $k'_{ir}, \ A$ and $B$ are taken as
$k'_{ir}=0$,
\begin{eqnarray}
A=B+\frac{ic'}{2k_{ir}}\phi_{ir}^{(p)}(0), \
B=A-\frac{ic'}{2k_{ir}}\phi_{ir}^{(p)}(0)
\end{eqnarray}
with $k_{ir}=k_{ir}^{(0)}$, Eq. (9) becomes the well-known
solutions \cite{Ge2}, \cite{Mattis}
\begin{eqnarray}
\phi_{ir}^{(p)}&=&\theta (x_{ir})\Big[B e^{ik_{ir}x_{ir}} +A
e^{-ik_{ir}x_{ir}}\Big] \nonumber \\ &+&\theta (-x_{ir})\Big[A
e^{ik_{ir}x_{ir}} +Be^{-ik_{ir}x_{ir}}\Big],
\end{eqnarray}
which implies
\begin{eqnarray}
\phi_{ir}^{(p)}(0)=A+B.
\end{eqnarray}
The selection to $A$ and $B$ in Eq. (10) is because of the
requirement of exchange symmetric Eq. (11),
$\phi_{ir}^{(p)}(x_{ir})=\phi_{ir}^{(p)}(-x_{ir})$. Solving Eq.
(10) and Eq. (12) yields
\begin{eqnarray}
A=\frac 1 2 \Big(1+\frac{ic'}{k_{ir}}\Big)\phi_{ir}^{(p)}(0), \
B=\frac 1 2 \Big(1-\frac{ic'}{k_{ir}}\Big)\phi_{ir}^{(p)}(0).
\end{eqnarray}
Applying this to Eq. (11), the normalization of $\phi_{ir}^{(p)}$
gives $\phi_{ir}^{(p)}(0)\varpropto 1/\sqrt{L}$ with $L$ being the
length of motion region adjusted by the boundary conditions
\cite{Ge2}. Obviously, for the plane wave solution (11) the
average kinetic energy equates the total relative energy so that
the nonzero average potential leads to the paradox on energy
expectation value. Only for an infinite $L$ the average potential
$c|\phi (0)|^2$ vanishes and the contradiction does not exist. It
is the non-zero energy correction $E'_{ir}$ that helps us to
overcome the difficulty on energy expectation value for the case
of finite motion region.

When $k'_{ir}\ne 0$ is set, Eq. (9) cannot be exactly solved.
However, it allows existence of the infinite series solution in
terms of plane waves of different wave vectors. The series
expansion leads the terms proportional to $k'^2_{ir}$ to contain
the harmonic waves of various momentums, which are generated by
the pair interactions.

For $k_{ir}^{(0)}\gg k'_{ir}$ we can treat the terms proportional
to $k'^2_{ir}$ as perturbation and construct approximate solution
of the equation easily, through the unperturbed solution (11).
This requires $c|\phi_{ir}^{(p)}(0)|^2\sim E'_{ir} \ll
E_{ir}^{(0)}$, since $E'_{ir}$ is the corrected energy from the
average potential. Note that only for small $c$ or big $L$ (small
$\phi_{ir}^{(p)}(0)$), it is valid, because of the small
$E'_{ir}$. In fact, simply inserting Eq. (11) into left hand side
of Eq. (9) to replace $\phi_{ir}^{(p)}$, we immediately arrive at
the perturbed solution \cite{Hai}, \cite{Hai2}. Noticing
$k_{ir}\approx k^{(0)}_{ir}+k'^2_{ir}/(2k^{(0)}_{ir})$, for
simplicity we let $k_{ir}$ in denominator of Eq. (13) be equal to
$k_{ir}^{(0)}$ approximately. To avoid resonance, this
approximation does not be used for the exponential functions in
Eq. (11). The perturbed solution then reads as
\begin{widetext}
\begin{equation}
\phi_{ir}^{(p)}=\theta (x_{ir})\Big\{\Big[1
+e^{ik'^2_{ir}x_{ir}/(2k^{(0)}_{ir})}-e^{ik'^2_{ir}C/(2k^{(0)}_{ir})}
\Big]Be^{ik_{ir}^{(0)}x_{ir}} +\Big[1
+e^{-ik'^2_{ir}x_{ir}/(2k^{(0)}_{ir})}-e^{-ik'^2_{ir}D/(2k^{(0)}_{ir})}\Big]A
e^{-ik_{ir}^{(0)}x_{ir}}\Big\} \nonumber
\end{equation}
\begin{equation}
+ \theta (-x_{ir})\Big\{\Big[1
+e^{ik'^2_{ir}x_{ir}/(2k^{(0)}_{ir})}-e^{ik'^2_{ir}C/(2k^{(0)}_{ir})}
\Big]Ae^{ik_{ir}^{(0)}x_{ir}} +\Big[1
+e^{-ik'^2_{ir}x_{ir}/(2k^{(0)}_{ir})}-e^{-ik'^2_{ir}D/(2k^{(0)}_{ir})}\Big]B
e^{-ik_{ir}^{(0)}x_{ir}}\Big\}+O(k'^2_{ir}),
\end{equation}
\end{widetext}
where $A$ and $B$ obey Eq. (13) with $k_{ir}= k_{ir}^{(0)}$,
$O(k'^2_{ir})$ denotes the terms proportional to $k'^2_{ir}$. The
exchange symmetry of $\phi_{ir}^{(p)}$ infers
$e^{ik'^2_{ir}C/(2k^{(0)}_{ir})}=e^{-ik'^2_{ir}D/(2k^{(0)}_{ir})}$,
namely $-k'^2_{ir}(C+D)/(2k^{(0)}_{ir})=2n\pi$ for $n=1,2,\cdots$.
The terms containing $k'_{ir}$ give perturbed correction to
unperturbed solution (11) that includes two different waves with
wave vectors $k^{(0)}_{ir}$ and $k_{ir}$ respectively.

To simplify the discussions, we rewrite Eq. (14) as
$\phi_{ir}^{(p)}=\phi_{ir}^{(0)}+ \phi'_{ir}$ with
$\phi_{ir}^{(0)}=\phi_{ir}^{(p)}|_{k'_{ir}=0}$, and let $H_{ir}$
be the Hamiltonian associated with Eq. (6). Thus the constant
$\phi_{ir}^{(p)}(0)$ is determined by the normalization of
$\phi_{ir}^{(0)}$, while the relation among $k'_{ir}$,
$k_{ir}^{(0)}$ and $c$ can be give by $\langle\phi_{ir}^{(p)}|
H_{ir}|\phi_{ir}^{(p)}\rangle=E_{ir}^{(0)}+E'_{ir}$, no paradox of
the energy expectation value occurs. Therefore, the solution (14)
is physically correct one.  Noticing $\langle\phi_{ir}^{(0)}|
H_{ir}|\phi_{ir}^{(0)}\rangle=E_{ir}^{(0)}+c|\phi_{ir}^{(0)}(0)|^2/\sqrt{2}$,
we have the corrected energy
\begin{eqnarray}
E'_{ir}&=&\frac{\hbar^2
k'^2_{ir}}{2m}=\frac{c}{\sqrt{2}}|\phi_{ir}^{(0)}(0)|^2+\langle\phi'_{ir}|
H_{ir}|\phi_{ir}^{(0)}\rangle \nonumber
\\ &+&\langle\phi^{(0)}_{ir}|
H_{ir}| \phi'_{ir}\rangle+\langle\phi'_{ir}|
H_{ir}|\phi'_{ir}\rangle.
\end{eqnarray}
Clearly, only for infinite $L$, $\phi_{ir}^{(0)}(0)=0$, Eq. (15)
allows $E'_{ir}=0, k'_{ir}=0$ such that
$\phi_{ir}^{(p)}=\phi_{ir}^{(0)}$, Eq. (11) becomes exact solution
of the integral equation (9).

Given the physical solution (14), the scattering amplitude can be
approximately obtained. We apply $k'^2_{ir}\approx 0$ and Eq. (13)
to Eq. (14), resulting in the scattering amplitude
\begin{eqnarray}
S(k_{ir})=\frac A B =\frac{k_{ir}+ic'}{k_{ir}-ic'}\approx
\frac{k_{ir}^{(0)}+k'^2_{ir}/(2k^{(0)}_{ir})+ic'}{k_{ir}^{(0)}
+k'^2_{ir}/(2k^{(0)}_{ir})-ic'},
\end{eqnarray}
which is similar to the known result \cite{Ge2}-\cite{Yang}. This
similarity infers that we can directly extend this result of
two-particle problem to $N$-boson system.

It is impossible, of course, to seek perturbed solution of the
integral equation (9) for large average potential that
necessitates the mean-field approximation. After eliminating the
paradox on energy expectation value, the contradiction implied in
mean-field theory of 1D $N$-boson system will be avoided. In fact,
adopting Eqs. (2) and (9), the average Hamiltonian of second
quantization reads as $\langle\psi| H|\psi\rangle$ with $H$
including the kinetic energy and internal pair interaction, even
external potential. Differing from the plane wave solution, the
formally exact solution (9) does not identify the average kinetic
energy with the total average energy and allows existence of the
nonzero average potential thereby. Thus the well known mean-field
theory is valid for the zero-range pair interacting system that
means the wave function obeying a nonlinear Schr\"{o}dinger
equation \cite{Lieb}-\cite{Leggett}.

In summary, we have investigated the 1D quantum $N$-boson system
with infinite zero-range pair interaction. The paradox of energy
expectation value was found for the plane wave solution previously
reported in some articles. The integral equation (8) that is
completely on an equality with the corresponding Schr\"{o}dinger
equation was established and was applied to construct the formally
physical solution (9) without the paradox. The approximate
solution and energy correction from the nonzero average potential
were detailed, and the scattering amplitude, harmonic waves of
various momentums and mean-field theory of the system were
discussed simply.

The method and results of this work are useful for treating any
quantum system with the singular delta potential. For instance, in
the case of a delta potential well with $c<0$, the relative energy
$E_{ir}$ is negative and the "wave vector" $k_{ir}=i|k_{ir}|$ is
imaginary. Inserting these into Eqs. (8) and (9) and selecting
suitable constants $A, \ B, \ C$ and $D$ to satisfy the
boundedness conditions as in Ref. \cite{Hai}, we can directly
construct the bound state solution of the system with nonzero
energy correction and without any paradox. It is straightforward
to extrapolate the results to the 1D system with combined
potential of the pair interaction and harmonic trap, which has
often been employed to treat the Bose-Einstein condensates. It is
worth noting that in the bound states the integral equation (8)
has no physical exact solution even for infinite motion region
that necessitates approximate and numerical solutions. The quantum
many-body problem is a source of the quantum Yang-Baxter equation
\cite{Ge2} that is worth the further investigations.

\begin{acknowledgments}
This work was supported by the NNSF of China under Grant No.
10275023, and by the LMRAMP of China under Grant No. T152103.
\end{acknowledgments}

\end{document}